\documentclass[12pt]{iopart}

\usepackage{graphicx}
\usepackage{color}
\begin{document}

\title{Magnetic hexadecapole $\gamma $ transitions and neutrino-nuclear responses  
in medium-heavy nuclei}

\author[first]{Lotta Jokiniemi$^1$, Jouni Suhonen$^1$, and Hiroyasu Ejiri$^2$
}
\address{1. University of Jyvaskyla, Department of Physics, P.O.Box 35, FI-40014, 
University of Jyvaskyla, Finland \\ 
2. Research Center for Nuclear Physics, Osaka University, Osaka 567-0047, Japan 
}
\ead{lotta.m.jokiniemi@student.jyu.fi, jouni.suhonen@phys.jyu.fi, ejiri@rcnp.osaka-u.ac.jp}

\vspace{10pt}
\begin{indented}
\item[] 
\end{indented}

\begin{abstract}
Neutrino-nuclear responses in the form of squares of nuclear matrix elements,NMEs, 
are crucial for studies of neutrino-induced processes in nuclei. In this work we
investigate magnetic hexadecapole (M4) NMEs in medium-heavy nuclei.
The experimentally derived NMEs, $M_{\rm EXP}$(M4), deduced from  
observed  M4 $\gamma $ transition half-lives are compared with the
single-quasiparticle (QP) NMEs, $M_{\rm QP}$(M4), 
and the microscopic quasiparticle-phonon model (MQPM) NMEs 
$M_{\rm MQPM}$(M4). The experimentally driven M4 NMEs are found to 
be reduced by a coefficient $k \approx 0.29 $ with respect 
to $M_{\rm QP}$(M4) and by $k \approx 0.33$ with respect to $M_{\rm MQPM}$(M4). 
The M4 NMEs are reduced a little by the quasiparticle-phonon correlations of the MQPM 
wave functions but mainly by other nucleonic and non-nucleonic correlations 
which are not explicitly included in the MQPM.
The found reduction rates are of the same order of magnitude as those for
magnetic quadrupole $\gamma $ transitions and Gamow-Teller (GT) and spin-dipole (SD) 
$\beta $ transitions. The impact of the found reduction coefficients on the magnitudes of the NMEs 
involved in astroneutrino interactions and neutrinoless double beta decays
are discussed. \\

\end{abstract}


\section{Introduction}

Neutrino interactions in nuclei are studied e.g. by investigating scatterings of
astro-neutrinos on nuclei and by the attempts to record the neutrinoless double 
beta ($0\nu \beta \beta $) decays. Here the neutrino-nuclear responses can be condensed
in the squares of nuclear matrix elements (NMEs) and it is necessary to study through
them the neutrino properties and astro-neutrino reactions that are of interest to 
particle physics and astrophysics, as discussed in the review 
articles \cite{eji00,ver12,avi08,suh98} and references therein. 

The present work aims at investigating the magnetic hexadecapole (M4) $\gamma $ NMEs, 
$M_{\gamma}$(M4), in medium-heavy nuclei to study higher-multipole axial-vector NMEs 
associated with higher-energy components of astro-neutrino reactions and 
$0\nu \beta \beta $ decays. Such components are shown to be important for, e.g., the
$0\nu \beta \beta $ decays \cite{hyv15}.
 
Neutrino-nuclear responses associated with neutral-current (NC) and charged-current 
(CC) interactions are studied by investigating the relevant $\gamma $ and $\beta $ 
decay transitions
or NC and CC scatterings on nuclei. The momenta involved in astro-neutrino scatterings and 
$0\nu \beta \beta $ decays are of the order of $50-100$ MeV/c. Accordingly, depending
on the involved momentum exchanges, the multipoles $J^{\pi}$ with angular momenta $J$ up to 
around 4-5 are involved (e.g. $0\nu \beta \beta $ decays mediated by light Majorana neutrinos,
see \cite{hyv15}), or even higher multipoles can be engaged ($0\nu \beta \beta $ decays 
mediated by heavy Majorana neutrinos, see \cite{hyv15}).

In some previous works, axial-vector CC resonances of GT(1$^+$) and SD(2$^-$) NMEs for 
allowed and first-forbidden $\beta $ transitions are shown to be reduced much in 
comparison with the quasiparticle (QP) and pnQRPA (proton-neutron quasiparticle
random-phase approximation) NMEs \cite{fae08,suh13,eji14,suh14,eji15} 
due to spin-isospin ($\sigma \tau$) nucleonic and non-nucleonic correlations and 
nuclear medium effects. These studies show that
exact theoretical evaluations for the astro-neutrino and $0\nu \beta \beta $ NMEs, 
including a possible renormalization of the axial-vector coupling constant $g_{\rm A}$, 
are hard. The corresponding NC nuclear responses of magnetic dipole (M1) and quadrupole 
(M2) $\gamma $ transitions are also known to be much reduced with respect to the QP 
NMEs \cite{eji78}. Similar studies have been conducted in the case of the two-neutrino
double beta decays in \cite{bar13,bar15} in the framework of the IBA-2 model. Also
the derivation of effective operators has been proposed \cite{hol13}.
All these studies bear relevance to the previously mentioned 
Majorana-neutrino mediated $0\nu \beta \beta $ decays, to high-energy astro-neutrino
reactions, but also to the lower-energy (up to 30 MeV) supernova-neutrino scatterings 
off nuclei, as shown e.g. in \cite{vol02,alm13,alm15,laz07}.

In the light of the above discussions it is of great interest to investigate the
spin-hexadecapole (4$^{-}$) NMEs to see how the higher-multipole NMEs are reduced by the 
nucleonic and non-nucleonic spin-isospin correlations. Actually, there are almost no 
experimental CC hexadecapole $\beta $ NMEs in medium-heavy nuclei since
the $\beta $ decays are very rare third-forbidden unique transitions. However, it 
turns out that there are few measurements of the half-lives and electron spectra of 
the more complex fourth-forbidden non-unique $\beta$ transitions and they can serve
as potential testing grounds concerning the quenching effects of the weak vector 
($g_{\rm V}$) and axial-vector ($g_{\rm A}$) coupling constants \cite{haa16}.
On the other hand, there are many experimental data on NC M4 $\gamma $ NMEs, where 
the isovector component of the $\gamma $ NME is related to 
the analogous $\beta $ NME on the basis of the isospin symmetry. Thus we discuss mainly 
the M4 $ \gamma $ transitions in the present report in the aim to
help evaluate/confirm e.g. the $0\nu \beta \beta $ NMEs concerning their higher-multipole
aspects.

\section{Experimental M4 NMEs}

Here we discuss stretched M4 $\gamma $ transitions with $J_i=J_f \pm J$, where $J_i$
and $J_f$ are the initial and final state spins and $J$=4. 
The M4 $\gamma $ transition rate (per sec) is given in terms of the reduced M4 
$\gamma $ strength $B_{\gamma}$(M4) as \cite{boh75}
\begin{equation}
T(\textrm{M}4) = 1.87\times 10^{-6} E^9 B_{\gamma}(\textrm{M}4) (1+\alpha)^{-1},
\end{equation}
where  $E$ is the $\gamma $ ray energy in units of MeV, and $\alpha$ is the 
conversion-electron coefficient.
The reduced strength is expressed in terms of the M4 $\gamma $ NME in units of 
$e\hbar/(2Mc)$ fm$^3$ as
\begin{equation}
B_{\gamma}(\textrm{M}4) = (2J_i+1)^{-1} [M_{\gamma }(\textrm{M}4)]^2.
\end{equation}
 
The M4 $\gamma $ NME is expressed in terms of the M4  $\gamma $ coupling constants 
$g$(M4) and the M4 matrix element $M$(M4) as 
\begin{equation}
M_{\gamma} (\textrm{M}4) = g_{\rm p}(\textrm{M}4) \tau_{\rm p} M(\textrm{M}4) + 
g_{\rm n}(\textrm{M}4) \tau_{\rm n} M(\textrm{M}4), 
\label{eq:NME1}
\end{equation}
where the first and the second terms are for the odd-proton ($\tau_{\rm p}=(1-\tau_3)/2$) 
and odd-neutron ($\tau _{\rm n} = (1+\tau _ 3)/2$) transition NMEs with $\tau_3$ being the isospin $z$ component 
($\tau_3$=1 for neutron and $\tau_3=-1$ for proton). 
The $\gamma $ coupling constant is written as 
\begin{equation}
g_i(\textrm{M}4) = \frac{e\hbar}{2Mc}6(\mu_{i}-\frac{1}{5}g_{i}),
\label{eq:NME2}
\end{equation}
where $i={\rm p}$ for proton and $i={\rm n}$ for neutron, $\mu_{\rm p}=2.79$ 
and $\mu_{\rm n}=-1.91$ are the
proton and neutron magnetic moments and  $g_{l{\rm p}}=1$ and $g_{l{\rm n}}=0$ are
the proton and neutron orbital $g$ coefficients. The M4 $\gamma $ 
matrix element is expressed as 
\begin{equation}
M(\textrm{M}4) = <f||i^3 r^3 [\sigma \times Y_3]_4||i>,
\end{equation}
where $r$ is the nuclear radius and $Y_3$ is the spherical harmonic for multipole $l=3$.

\begin{table}[htpb] 
\caption{$M$(M4) NMEs for M4 $\gamma $ transitions
in the mass region of $A=70-120$, where the major  
single-QP transition is $1\textrm{g}_{9/2} - 2\textrm{p}_{1/2}$. 
Here p/n stands for the odd-proton/odd-neutron transition.
$M_{\rm EXP}, M_{\rm QP}$ and $M_{\rm MQPM}$ are the experimental, 
single-QP, and MQPM NMEs in units of 10$^3$ fm$^3$ . \label{tab:1}}
\centering
\begin{tabular}{cccccc}
\br\\
Nucleus & transition  & p/n & $M_{\rm EXP}$ & $M_{\rm QP}$ & $M_{\rm MQPM}$ \\
\mr
$~~^{85}$Kr  &  $2{\rm p}_{1/2} \rightarrow 1{\rm g}_{9/2}$ & n & 0.528 & 1.57 & 1.44 \\
$~~^{89}$Y & $2{\rm p}_{1/2} \rightarrow 1{\rm g}_{9/2}$ & p & 0.739& 2.12& 2.02\\
$~~^{89}$Zr  & $2{\rm p}_{1/2} \rightarrow 1{\rm g}_{9/2}$ & n  & 0.559 &  1.60 & 1.47 \\
$~~^{91}$Y  &  $1{\rm g}_{9/2} \rightarrow 2{\rm p}_{1/2}$ & p & 0.480 & 2.13& 1.83\\
$~~^{105}$In  & $2{\rm p}_{1/2} \rightarrow 1{\rm g}_{9/2}$ & p  & 0.706 &  2.38 & 2.09\\
$~~^{107}$In  & $2{\rm p}_{1/2} \rightarrow 1{\rm g}_{9/2}$ & p  & 0.670 &  2.40 & 2.13 \\
$~~^{109}$In  & $2{\rm p}_{1/2} \rightarrow 1{\rm g}_{9/2}$ & p  & 0.640 &  2.33 & 2.10 \\
$~~^{111}$In  & $2{\rm p}_{1/2} \rightarrow 1{\rm g}_{9/2}$ & p  & 0.609 &  2.45 & 2.03 \\
$~~^{113}$In  & $2{\rm p}_{1/2} \rightarrow 1{\rm g}_{9/2}$ & p  & 0.603 &  2.46 & 2.05\\
$~~^{115}$In  & $2{\rm p}_{1/2} \rightarrow 1{\rm g}_{9/2}$ & p  & 0.614&  2.48 & 2.03 \\
\br
\end{tabular}
\end{table}

\begin{table}[htb]
\caption{The same as Table~\ref{tab:1} for $A=130 - 150$, where 
the major single-quasiparticle transition is 
$1\textrm{h}_{11/2}-2\textrm{d}_{3/2}$. \label{tab:2}}
\centering
\begin{tabular}{cccccc}
\br \\
Nucleus & Transition  & p/n& $M_{\rm EXP}$ & $M_{\rm QP}$ & $M_{\rm MQPM}$ \\
\mr
$~~^{135}$Xe  & $1{\rm h}_{11/2} \rightarrow 2{\rm d}_{3/2}$ & n  & 1.11 &  3.12 & 2.87 \\
$~~^{137}$Ba  & $1{\rm h}_{11/2} \rightarrow 2{\rm d}_{3/2}$ & n  & 1.03 &  3.13 & 2.77\\
$~~^{139}$Ba  & $1{\rm h}_{11/2} \rightarrow 2{\rm d}_{3/2}$ & n  & 0.968 &  3.12 & 2.82 \\
$~~^{141}$Nd  & $1{\rm h}_{11/2} \rightarrow 2{\rm d}_{3/2}$ & n  & 0.893 &  3.17 & 2.79 \\
$~~^{143}$Sm  & $1{\rm h}_{11/2} \rightarrow 2{\rm d}_{3/2}$ & n  & 0.878&  3.19 & 2.81 \\
\br
\end{tabular}
\end{table}

\begin{figure}[htb]
\caption{
 Top: EXP: experimental NMEs $M_{\rm EXP}$(M4) for odd-neutron transitions (light-blue squares) and 
 odd-proton transitions (dark-blue diamonds). \\
 QP: quasiparticle NMEs $M_{\rm QP}$(M4) for odd-neutron transitions 
(light-blue tip-up triangles) and  
 odd-proton transitions (dark-blue tip-down triangles). \\
 Bottom: EXP: experimental NMEs $M_{\rm EXP}$(M4) for odd-neutron transitions 
(light-blue squares) and odd-proton transitions (dark-blue diamonds).\\
 MQPM: NMEs $M_{\rm MQPM}$(M4) for odd-neutron transitions (light-blue tip-up triangles) and  
 odd-proton transitions (dark-blue tip-down triangles). 
\label{fig:nme}}
\begin{center}
\includegraphics[width=0.6\textwidth]{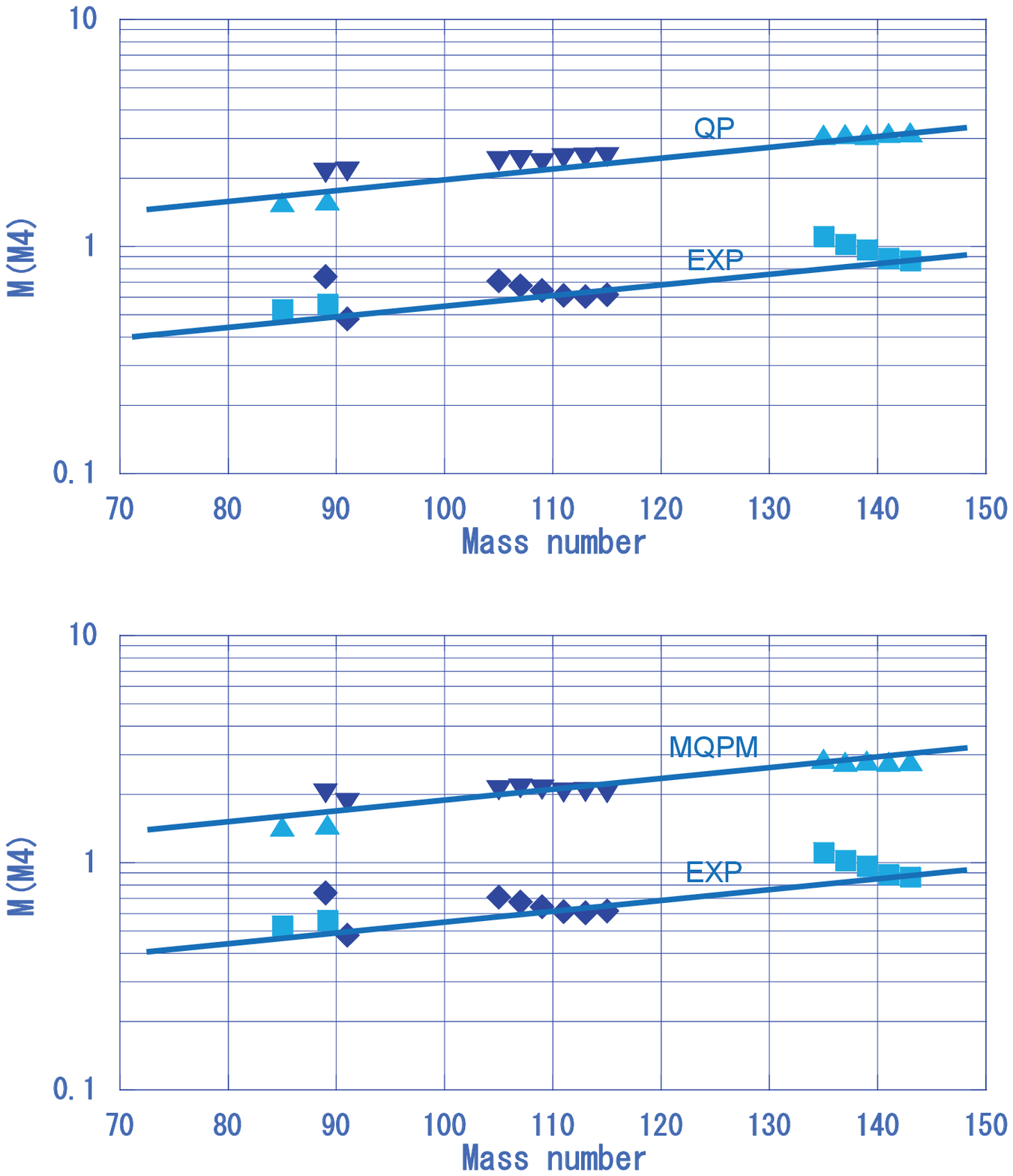}
\end{center}
\end{figure}

The isotopes used for on-going and/or future $\beta \beta $ experiments are  
$^{76}$Ge, $^{82}$Se, $^{96}$Zr, $^{100}$Mo, $^{116}$Cd, $^{130}$Te and $^{136}$Xe 
\cite{ver12}. They are in the mass regions of $A = 70-120$ and 
$A= 130-150$. The single-quasiparticle (single-QP) M4 transitions in these mass 
regions are uniquely tagged by the pairs $1\textrm{g}_{9/2}-2\textrm{p}_{1/2}$ and
$1\textrm{h}_{11/2}-2\textrm{d}_{3/2}$, respectively. 
Here the higher spin state is the intruder one from the higher major 
shell with opposite parity. The single-particle M4 NMEs corresponding to these tagging
transitions are quite large because of the large radial and angular overlap integrals. 

The single-quasiparticle M4 $\gamma $ transitions in the mentioned two mass 
regions are analyzed in Tables~\ref{tab:1} and \ref{tab:2}. The M4 NMEs derived 
from the experimental half-lives are given in the third column of these tables.

The values of $M_{\rm EXP}$(M4) are plotted against the mass number in 
Fig.~\ref{fig:nme}. They are around $(0.6\pm0.1)\times 10^{3}$ fm$^3$ and
$(1.0\pm 0.1)\times 10^{3}$ fm$^3$ for the two mass regions, respectively. 
They are well expressed as 
\begin{equation}
M_{\rm EXP}(\textrm{M}4) \approx 6 \times A  ~ {\rm fm}^3\,,
\end{equation}
where the mass number $A$ reflects the $r^3$ dependence of the M4 NME.


\section {Quasiparticle M4 NMEs}

The M4 $\gamma$ transitions given in Tables~\ref{tab:1} 
and \ref{tab:2} are all, in their simplest description, transitions between
single-quasiparticle states. The NMEs for the single-quasiparticle 
transitions are written by using the single-particle matrix element 
$M_{\rm SP}$(M4) and the pairing coefficient $P$ as   
\begin{equation}
M_{\rm QP}(\textrm{M4})= M_{\rm SP}(\textrm{M4})\,P_{ij} \,,
\label{eq:pairing}
\end{equation}
where the pairing coefficient is given by
\begin{equation}
P_{ij}=U_iU_f + V_iV_f \,,
\label{eq:pcoefficient}
\end{equation}
and $U_i$ ($U_f)$ and $V_i$ ($V_f)$ are the vacancy and occupation 
amplitudes for the initial  (final) state. The single-quasiparticle states discussed here 
are low-lying states located at the diffused Fermi surface, as shown in Fig. 2. 
Thus the occupation and vacancy probabilities are in the region of 
$U^2=1-V^2=0.5 \pm 0.3$, and the pairing coefficient is given roughly as $P\approx 1 $. 
In this work the single-quasiparticle NMEs $M_{\rm QP}$(M4) are calculated by using 
the BCS wave functions with H.O. single-particle states.  They are given in the 
fifth column of Tables~\ref{tab:1} and \ref{tab:2}, and are plotted in the top
panel of Fig. 1.

\begin{figure}[htb]
\caption{Schematic diagram of the energy (E) and the occupation probabilities, $V_i ^2$ and $V_f^2$, for 
the initial and final states (see body of text). The energy levels are shown by the horizontal lines.
Vacancy probabilities are given as $U_i^2=V_i^2-1$ and $U_f^2=V_f^2-1$.  
The paring coefficient $P_{ij}$ for the $\gamma $ transition is given by $U_iU_f+V_iV_f$.
\label{fig:strengths}}
\begin{center}
\includegraphics[width=0.35\textwidth]{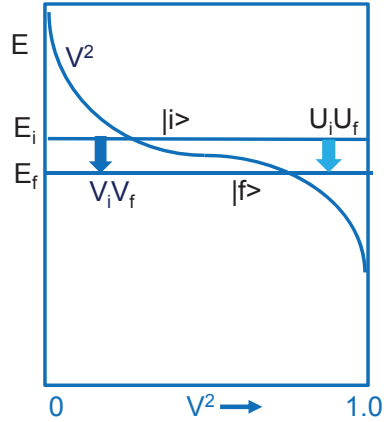}
\end{center}
\end{figure}

The experimental M4 matrix elements $M_{\rm EXP}$(M4) are uniformly smaller 
by a coefficient of around 0.29 $\pm$0.05 than the single-quasiparticle ones, $M_{\rm QP}$(M4). 
We introduce a reduction coefficient $k_ i$ as in the case of GT 1$^+$ and SD 2$^-$ 
\cite{eji14,eji15} transitions. It is defined as 
\begin{equation}
M_{\rm EXP}(\textrm{M}4)=k_i(M4)\,M_{\rm QP}(\textrm{M}4)\,, 
\label{eq:prop}
 \end{equation}
where $k_i$, with $i=$p,n, are the reduction coefficients for single quasi-proton 
and quasi-neutron 
M4 $\gamma $ transitions, respectively. The ratios $k_i(M4)$ are 
$k_{\rm p} \approx 0.3 $ and $k_{\rm n} \approx 0.3$, as shown in the top panel of Fig. 3. 
The found reductions are consistent with the reductions discussed in \cite{eji78}.

\begin{figure}[htb]
\caption{Reduction coefficients $k$ for the M4 NMEs.\\
 Top: $k$: ratios of experimental NMEs $M_{\rm EXP}$(M4) to the quasiparticle NMEs 
$M_{\rm QP}$(M4) for odd-neutron transitions (light-blue squares) and 
 odd-proton transitions (dark-blue diamonds).\\
 Bottom:  $k$: ratios of experimental NMEs $M_{\rm EXP}$(M4) to the MQPM NMEs 
$M_{\rm MQPM}$(M4) for odd-neutron transitions (light-blue squares) and 
 for odd-proton transitions (dark-blue diamonds). 
 \label{fig:reduction}}
\begin{center}
\includegraphics[width=0.6\textwidth]{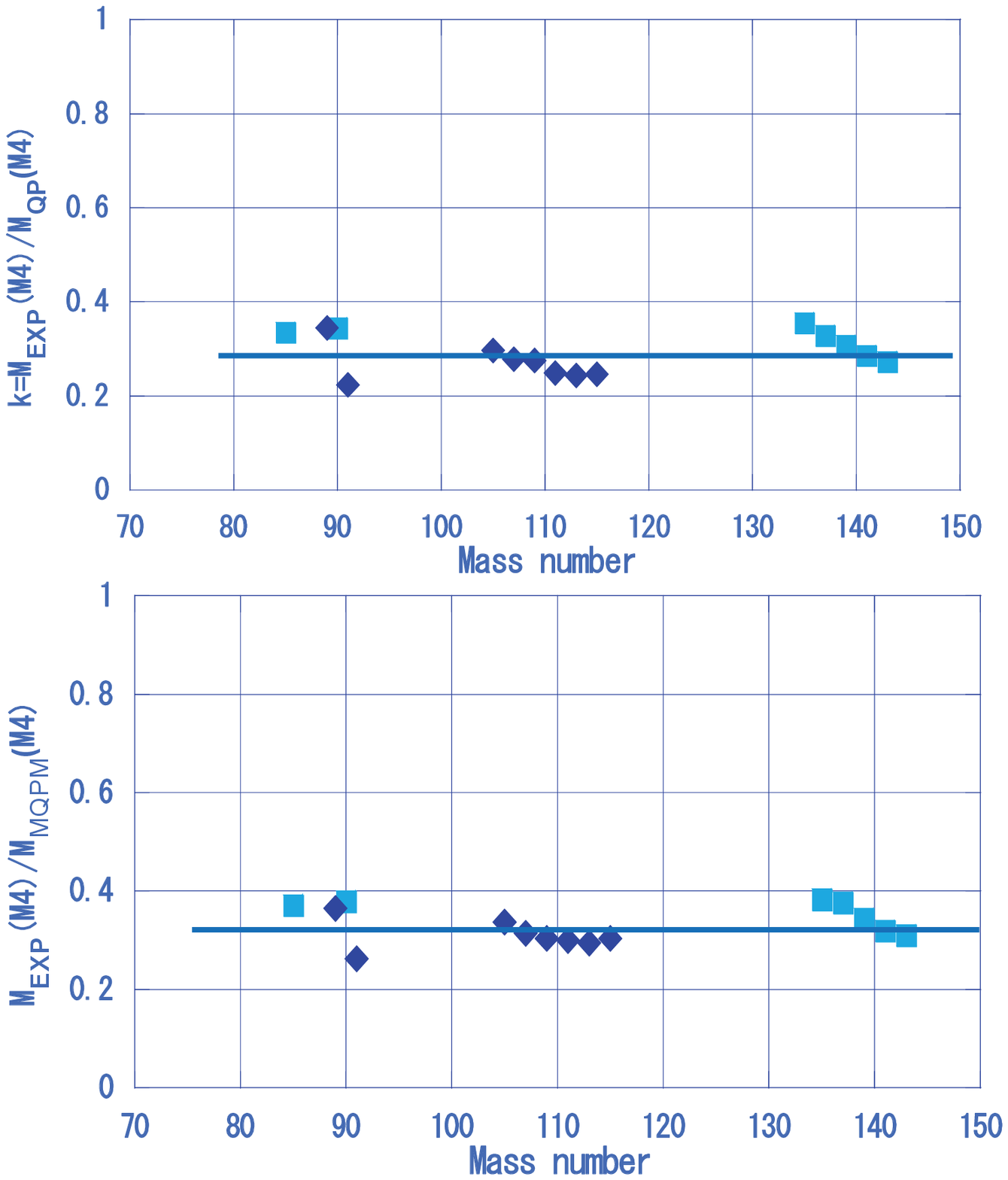}
\end{center}
\end{figure}


%
The quasiparticle NMEs $M_{\rm QP}$(M4) are calculated by assuming a stretched M4 transition
between the initial and final nuclear states (see column 2 of Tables~\ref{tab:1} 
and \ref{tab:2}) that are assumed to have a one-quasiparticle structure. These states
are thus described as
\begin{equation}
\vert\alpha\rangle = a_{\alpha}^{\dagger}\vert\textrm{BCS}\rangle \,,
\label{eq:qpstate}
\end{equation}
where $a_{\alpha}^{\dagger}$ creates a quasiparticle on a nuclear mean-field orbital with
quantum numbers $\alpha =a,m_{\alpha}$, where $a$ contains the principal quantum 
number $n$, the orbital angular
momentum ($l$) and total angular momentum ($j$) quantum numbers in the form $nl_j$ as
displayed in column 2 of Tables~\ref{tab:1} and \ref{tab:2}. Here
$m_{\alpha}$ is the $z$ projection of the total angular momentum and
$\vert\textrm{BCS}\rangle$ is the BCS vacuum. The quasiparticles are 
defined by the Bogolyubov-Valatin transformation as 
\begin{equation}\label{Eq:BV-trans}
\left\lbrace \begin{array}{l}  
a_{\alpha}^{\dagger}=  U_a c_{\alpha}^{\dagger}+V_a \tilde{c}_{\alpha} \,, \\
\tilde{a}_{\alpha}=  U_a \tilde{c}_{\alpha}-V_a c_{\alpha}^{\dagger} \,,
\end{array} \right.
\end{equation}  
where $c_{\alpha}^{\dagger}$ is the particle 
creation operator and the time-reversed particle annihilation operator 
$\tilde{c}_{\alpha}$ is defined by $\tilde{c}_{\alpha}=(-1)^{j_a+m_{\alpha}}c_{-\alpha}$ 
with $-\alpha=(a,-m_{\alpha})$. The $U$ and $V$ coefficients are the vacancy and occupation
amplitudes present also in the quasiparticle matrix elements of 
Eqs.~(\ref{eq:pairing}) and (\ref{eq:pcoefficient}). The values of these amplitudes are
obtained in BCS calculations (for details see, e.g., \cite{suh07}) and they are 
used also in the subsequent nuclear-structure
calculations of the odd-mass nuclei and their neighboring even-even-mass 
reference nuclei.

\section{Microscopic quasiparticle-phonon model for M4 NMEs} 

The microscopic quasiparticle-phonon model (MQPM) takes the structure of the nuclear states
beyond the approximation (\ref{eq:qpstate}). In the MQPM this extension is done in the
traditional way of starting from an even-even reference nucleus where the states are
described as QRPA (quasiparticle random-phase approximation) states called here phonons
since the lowest ones are usually collective vibrational states. These states can be
formally written as
\begin{equation}
\vert\omega\rangle = Q_{\omega}^{\dagger}\vert\textrm{QRPA}\rangle \,,
\label{eq:QRPA-state}
\end{equation}
where the phonon operator $Q_{\omega}^{\dagger}$ creates a nuclear state with 
quantum numbers $\omega$, containing
the angular momentum $J_{\omega}$, parity $\pi_{\omega}$ and the quantum number
$k_{\omega}$ which enumerates states with the same angular momentum and parity. 
The state (\ref{eq:QRPA-state}) is a linear combination of two-quasiparticle states as 
explicitly written in \cite{toi98} where the MQPM was first introduced. To arrive
at a state in the neighboring odd-mass nucleus one has to couple a proton (proton-odd
nucleus) or a neutron (neutron-odd nucleus) quasiparticle to the phonon operator
$Q_{\omega}^{\dagger}$ which is a two-quasiparticle operator. In this way one creates 
three-quasiparticle states in the traditional quasiparticle-phonon coupling scheme
and these states are then mixed with the one-quasiparticle states by the residual
nuclear Hamiltonian (for details see \cite{toi98}). Hence we obtain the MQPM states
\begin{equation}
\vert k\,j\,m\rangle = \Gamma_{k}^{\dagger}(jm)\vert\textrm{QRPA}\rangle \,,
\label{eq:MQPM-state}
\end{equation}
where a $k$:th state of angular momentum $j$ and its $z$ projection $m$ is created 
in an odd-mass nucleus by a creation operator which mixes one-quasiparticle and 
three-quasiparticle components in the form
\begin{equation}\label{eq:MQPM-op}
\Gamma_k^{\dagger}(jm)=\sum_n X_{n}^{k} a_{n j m}^{\dagger}+
\sum_{a \omega}X^{k}_{a\omega}[a_a^{\dagger}Q_{\omega}^{\dagger}]_{jm}\,,
\end{equation}
where the first term is the one-quasiparticle contribution and the second term is the
quasiparticle-phonon contribution. The amplitudes 
$X^k_{n}$ and $X^k_{a\omega}$ are computed from the MQPM equations of motion 
\cite{toi98}. In solving these equations special care is to be taken to 
handle the over-completeness and the non-orthogonality of the 
quasiparticle-phonon basis, as described in detail in \cite{toi98}.     

In the actual calculations we used slightly modified Woods-Saxon single-particle energies 
to improve the quality of the computed energy spectra of the odd-mass nuclei
involved in the present work. This
resulted in a good correspondence between the computed and experimental low-energy spectra
of these nuclei. We adopted a residual Hamiltonian with realistic effective two-nucleon 
interactions derived from the Bonn-A one-boson-exchange potential \cite{hol81}. 
The free parameters of the interaction were fixed in the BCS and QRPA phases of the 
calculations as explained in \cite{suh88}. The two-body monopole matrix elements were
multiplied by one parameter for protons and one for neutrons to scale phenomenologically
the proton and neutron pairing strengths separately. This was done by fitting the
computed pairing gaps to the phenomenological ones, derived from the measured proton
and neutron separation energies \cite{aud12}. The QRPA step contained two parameters
for each multipole $J^{\pi}$ to control the energies of the even-even excited states. These
were the strengths of the particle-hole and particle-particle parts of the two-nucleon
interaction. The particle-hole interaction controls the energies of collective states
and thus it was fitted to reproduce the experimental excitation energy of the lowest
state of a given multipolarity $J^{\pi}$, whenever data existed. When no data was
available the bare G-matrix was used in the calculations. Also the particle-particle
part of the multipole interaction was kept as bare G-matrix interaction.

After performing the BCS and QRPA 
calculations in the reference even-even nuclei, the initial and final nuclear states
in the neighboring odd-mass nuclei, of interest in the present work, were formed by
first creating the quasiparticle-phonon components of the operator (\ref{eq:MQPM-op}).
The convergence of the MQPM results for the energies of the involved states and the M4
transition amplitudes between them were monitored by adding more and more QRPA phonons
of different multipolarities $\omega$ in the diagonalization of the residual Hamiltonian.
In terms of the cut-off energy of the added phonons the convergence was achieved at around
14 MeV.

The converged M4 results are given in the 
sixth column of Tables~\ref{tab:1} and \ref{tab:2}, and are plotted in Fig. 1, bottom panel.
It is seen that the experimental M4 matrix elements, $M_{\rm QP}$(M4), are uniformly smaller, 
by a coefficient around 0.33, than the MQPM NMEs $M_{\rm MQPM}$(M4). 
Actually, the MQPM NMEs $M_{\rm MQPM}$(M4)) are 10-20 $\%$ smaller than the QP NMEs 
$M_{\rm QP}$(M4), as shown in Fig. 4. The admixtures of the quasiparticle-phonon 
components in the wave functions (\ref{eq:MQPM-op}) of the M4 initial and final 
states reduce the M4 NMEs a little, but not nearly enough to bring the MQPM NMEs close
to the corresponding experimental ones, at least by using the bare $g$ coefficients
(\ref{eq:NME2}) adopted in the present work. Hence, the major part of reduction from 
the MQPM NME to the experimental one (a reduction coefficient of 0.3) is considered to 
be due to such nucleonic and non-nucleonic $\tau \sigma$ correlations and nuclear-medium 
effects that are not explicitly included in the (traditional) quasiparticle-phonon 
coupling scheme that MQPM uses, i.e. the even-even nucleus serving as a reference for 
the odd-mass one.

\begin{figure}[htb]
\caption{Ratios of the quasiparticle NMEs $M_{\rm QP}$(M4) to the MQPM NMEs 
$M_{\rm MQPM}$(M4) for odd-neutron transitions (light-blue squares) and 
odd-proton transitions (dark-blue diamonds).
\label{fig:NME ratio}}
\begin{center}
\includegraphics[width=0.6\textwidth]{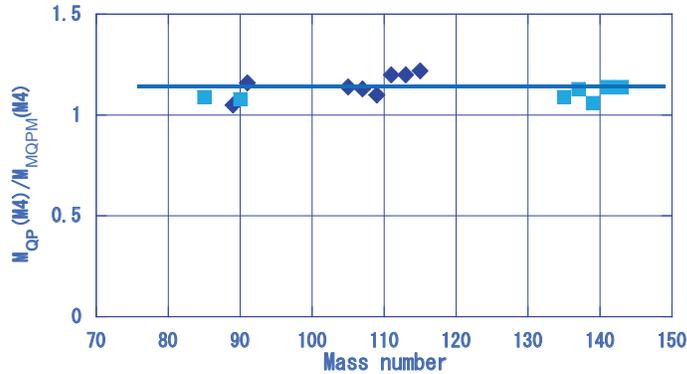}
\end{center}
\end{figure}

\section{Reduction of the axial-vector NMEs}

The $\gamma $ NME is decomposed into the isovector and isoscalar ones as \cite{eji78}
\begin{equation}
M_{\gamma} (\textrm{M}4) = g_{-}(\textrm{M}4)\frac{\tau_3}{2}M(\textrm{M}4) +  
g_{+}(\textrm{M}4)\frac{\tau_0}{2}M(\textrm{M}4) \,, 
\label{eq:iso-NME1}
\end{equation}
where  $M_{\gamma} (\textrm{M}4)$ is $g_{\rm p}M(\textrm{M}4)$ and $g_{\rm n}M(\textrm{M}4)$ 
for an odd proton and odd neutron transition, respectively, and
$g_{-}$ and $g_{+}$  are the isovector and isoscalar $\gamma $ coupling constants. 
They are written as 
\begin{equation}
g_{\pm}(\textrm{M}4) = \frac{e\hbar}{2Mc}6(\mu_{\pm}-\frac{1}{5}g_{l\pm}) \,,
\label{eq:iso-NME2}
\end{equation}
with $\mu _{\pm} = \mu _{\rm n} \pm\mu_{\rm p}$ and 
$g_{l\pm} =g_{l{\rm n}}\pm g_{l{\rm p}} \approx  \mp 1$.

The proton, neutron, isovector and isoscalar M4 $\gamma $ coupling constants are given as
\begin{equation}
g_{\rm p}=2.59 G, ~~g_{\rm n}=-1.91 G \,,
\end{equation}
\begin{equation}
g_-=-4.50 G, ~~g_+=0.68 G \,,
\end{equation}
where $G=6e\hbar(2Mc)^{-1}$ is the M4 $\gamma $ coupling coefficient.

The M4 NMEs in Eq. (\ref{eq:iso-NME1}) are rewritten as $k_ig_iM_{\rm QP}(\textrm{M}4)$ 
with $i=p,n,-,+ $, the symbols standing for the proton, neutron, isovector and 
isoscalar components, respectively. Then the reduction coefficients  and the weak 
couplings for proton and neutron transitions are  expressed in terms of those for 
the isovector and isoscalar components as \cite{eji78} 
\begin{equation}
k_{\rm p}g_{\rm p}= -\frac{g_-}{2}k_- + \frac{g_+}{2}k_+ \,,
\end{equation}
\begin{equation}
k_{\rm n} g_{\rm n}= \frac{g_-}{2}k_- + \frac{g_+}{2}k_+ \,,
\end{equation}
After this the isovector and isoscalar reduction coefficients can be derived as 
\begin{equation}
k_-=k_{\rm n} - 0.567 (k_{\rm n} - k_{\rm p}) \,,
\label{eq:kminus}
\end{equation}
\begin{equation}
k_+=k_{\rm n} - 3.89 (k_{\rm n} - k_{\rm p}) \,.
\label{eq:kplus}
\end{equation}
Since the experimentally derived NMEs for the proton and neutron transitions are 
approximately the same, i.e. $k_{\rm n} \approx k_{\rm p}$,  we get from Eq. (\ref{eq:kminus})  
$k_- \approx  k_{\rm n} \approx k_{\rm p} \approx 0.3$.
The isovector M4 NMEs are reduced by the coefficient $k_{-} \approx $ 0.3, in the same way
as the M4 $\gamma $ transition NMEs. It is notable that the amount of reduction 
for the isovector 4$^-$ $\gamma $ NMEs is the same as that found for the GT 1$^+$ 
and SD 2$^-$ $\beta$-decay NMEs \cite{eji14,eji15}.


Axial-vector $\beta $ and $\gamma $ NMEs for low-lying states are much reduced with 
respect to the QP NMEs due to the strong repulsive $\tau \sigma $ interactions 
since the axial-vector $\tau \sigma $ strengths are
pushed up into the $\tau \sigma$ giant-resonance (GR) and the $\Delta $-isobar regions. 
The M4 reduction coefficient of $k$(M4) $\approx 0.29 $ is nearly the same as the 
coefficient $k$(M2) $\approx 0.24$ for M2 $\gamma $ 
transitions and $k$(GT) $\approx 0.235$ and $k$(SD) 
$\approx 0.18 $  in the GT 1$^+$ and SD 2$^-$ $\beta$-decay NMEs, 
respectively \cite{eji14,eji15}. These reduction coefficients are plotted in Fig. 5 with
$\lambda$ denoting the angular-momentum content of the transition operator.

\begin{figure}[h]
\caption{Reduction coefficients $k(\textrm{M}) = M_{\rm EXP}(\textrm{M})/M_{\rm QP}(\textrm{M})$  
for the multipolarity M$=$M2 ($\lambda=2$) $\gamma$ and M$=$M4 ($\lambda=4$) $\gamma $ 
transitions (dark-blue diamonds) and those for M$=$GT ($\lambda=1$) $\beta $ and 
M$=$SD ($\lambda=2$) $\beta $ transitions (light-blue squares) in medium-heavy nuclei. 
Here $\lambda$ denotes the transition angular momentum.  The dotted line shows 
$k(\textrm{M})$=0.27. \label{fig:reduction2}}
\begin{center}
\includegraphics[width=0.5\textwidth]{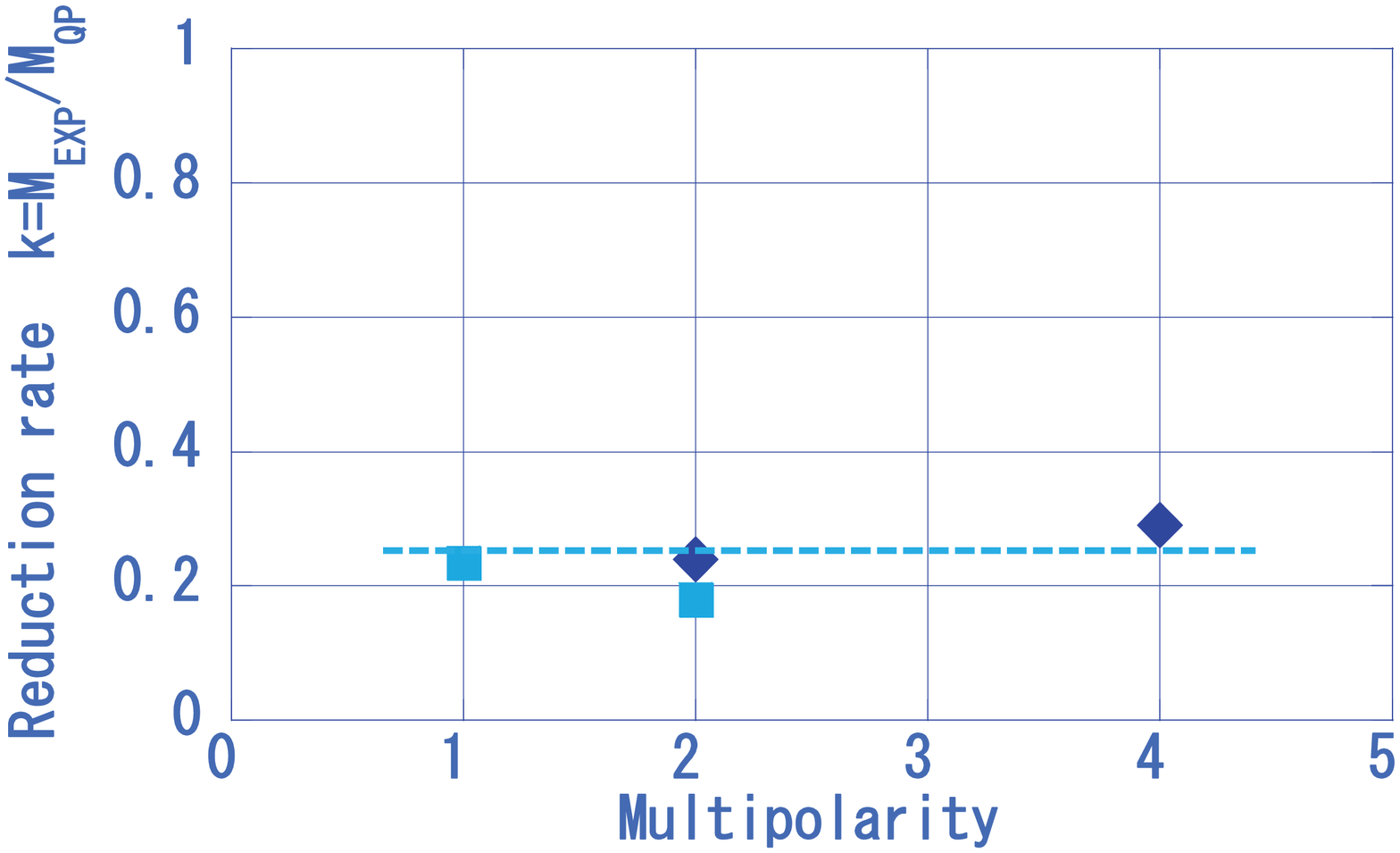}
\end{center}
\end{figure}

It seems that the reduction coefficients of the axial-vector NMEs are universal for NC 
and CC NMEs and for the angular momenta of $\lambda=1-4$. The reduction is considered 
to be due to such $\sigma \tau$ polarization interactions and nuclear-medium 
effects that are not explicitly included in the models. Then the reduction rate 
$k(\textrm{M})\approx 0.2-0.3$, with 
respect to the QP NME, is expressed as \cite{eji14,eji15} 
\begin{equation}
k(\textrm{M})=k_{\sigma \tau}(\textrm{M}) \times k_{\rm NM}(\textrm{M}),
\end{equation}
where  $k_{\sigma \tau}(\textrm{M})\approx 0.5$ and  $k_{\rm NM}(\textrm{M})\approx 0.5$ 
stand for the reductions due to the nucleonic $\sigma \tau$ polarization effects
and non-nucleonic $\Delta$-isobar and nuclear-medium effects, respectively. 

These universal effects may be represented by  the effective weak coupling 
$g_{\rm A}^{\rm eff}$, depending on the model NME, to incorporate the effects that 
are not explicitly included in the nuclear-structure model. In case of the pnQRPA 
with explicit nucleonic $\sigma \tau$ correlations, one may use $g_{\rm A}\approx 0.6$,
while in the case of the QP model, without any $\sigma \tau$ correlations, 
$g_{\rm A} \approx 0.3$, both in units of the bare value of $g_{\rm A}=1.26$ $g_V$. 

Here we note that the $\tau \sigma$ repulsive interaction concentrates the 
$\tau \sigma$ strength to the  highly excited $\tau \sigma $ GR, resulting in the 
reduction of the $\tau\sigma $ NMEs for low-lying states. Such reduction effect is 
incorporated in the pnQRPA with the strong $\tau \sigma$ interaction, so that the 
reduction factor is improved from $k_{\rm QP}\approx 0.3$ to 
$k_{\rm QRPA}\approx 0.6$. On the other hand, such strong $\tau \sigma$ interaction 
to give rise to the possible M4 GR is not explicitly incorporated in the MQPM, and thus
the reduction factor is only a little improved from $k_{\rm QP}\approx 0.29$
to $k_{\rm MQPM}\approx 0.33$.

\section{Discussion and conclusions}

Among other models, the QRPA-based models are used to compute the NMEs of double beta
decays \cite{suh98,suh12a}. In these calculations the importance of the quenching
of $g_{\rm A}$ magnifies since the 0$\nu \beta \beta $ NME includes the axial-vector 
NME proportional to the square of  $g_{\rm A}$. 
The M4 NC $\gamma$ results of the present investigation, together with the
earlier M2 NC and GT and SD CC results suggest that many of the leading multipoles
in the decomposition of the neutrinoless $\beta\beta$ NMEs are quenched roughly by 
the same amount by the $\sigma\tau$ correlations of the $\Delta$ region as well 
as other nuclear medium effects. Since the corresponding investigations have 
been done using data on low-lying nuclear states it is safe to say that the 
quenching applies at least to the low-lying states in nuclei. These observations 
are very relevant for the two-neutrino $\beta\beta$ decays since they are low-energy
phenomena and usually involve only one or few lowest states in the intermediate
nucleus \cite{eji00,bha98,civ99}. 
A consistent description of these decays can be 
achieved by using the quenched $g_{\rm A}$ derived from the GT $\beta$ decays \cite{pir15}.

Actually, the observed single $\beta $ GT NMEs are reduced by the effective coupling 
constant $k^{\rm eff}$ with respect to the model NME, 
and thus  the observed 2$\nu\beta \beta $ NMEs are well reproduced by using the experimental 
$k^{\rm eff}$, i.e. the  experimental single $\beta $ NMEs for low-lying states 
\cite{eji00,ver12,eji09}. However, it should be kept in mind that these results, important
as such, cannot be directly applied to the $0\nu\beta\beta$ processes since there large
momentum exchanges are involved and also vector-type of NMEs and higher excited 
states contribute. In this case it is 
a big challenge to develop such nuclear models for the axial-vector weak processes  
that include explicitly appropriate nucleonic and non-nucleonic correlations. If
successful, then in such models one could use the axial weak coupling of 
 $g_{\rm A}=1.26g_{\rm V}$ and
be free from the uncertainties introduced by the effective (quenched) $g_{\rm A}^{\rm eff}$.

Neutrino-nucleus scatterings are important to probe many astrophysical phenomena, like
the solar and supernova neutrinos \cite{eji00,woo90,lan03}. The GT NMEs bring in
most of the contributions for solar neutrinos and low-energy supernova neutrinos for
neutrino energies below 15 MeV (see, e.g. \cite{alm13,alm15}). The SD NMEs play a
role for medium-energy neutrinos above 15 MeV. For the low-energy solar neutrinos
reliable calculations of the GT NMEs are needed to evaluate the SNU values for the
pp, $^7$Be, CNO and other neutrinos. Experimental GT strengths can also be used, if
available \cite{eji00}. The SD contributions can be important for the CC supernova
anti-neutrino scatterings off nuclei even at low energies, as shown in
\cite{vol02,alm13,alm15,laz07}.

The supernova-neutrino nucleosyntheses are sensitive to the neutrino CC and NC 
interactions, as discussed in a review article \cite{lan03}.
Here SD and higher-multipole NMEs are involved in the high-energy components of the 
supernova neutrinos. It is important for accurate evaluations of the isotope 
distributions to use appropriate NMEs and
effective weak couplings of $g_{\rm A}^{\rm eff}$ and $g_{\rm V}^{\rm eff}$. 
QRPA calculations were made for $^{92}$Nb nuclei in \cite{che12}.     

The involved GT and SD NMEs can be studied via beta decays in nuclei
where beta-decay data is available. In some of these studies a strong quenching of both
$g_{\rm A}$ and $g_{\rm V}$ has been conjectured \cite{war90,war91,suz12,zhi13}. Such
quenching for the higher multipoles is extremely hard to study, the present study
being a rather unique one in this respect. Quenching of the higher multipoles can
also be studied via high-forbidden beta decays \cite{haa16} but the available data
is extremely scarce at the moment. Perspectives for the studies of the quenching
of both $g_{\rm A}$ and $g_{\rm V}$ is given by the spectrum-shape method introduced in
\cite{haa16}. There the shape of the beta spectrum of the high-forbidden non-unique
beta decays has been studied for the determination of the possible quenching of the
weak constants. Use of this method can be boosted by future high-sensitive measurements
of electron spectra in underground laboratories.

Finally, it is worth pointing out that the universal reduction/quenching of the 
$\tau\sigma$ NME, including $g_{\rm A}$, is  related to the shift of the strength to 
the higher GR and $\Delta $ isobar regions. Charge-exchange reactions report about a
50-60 $\%$ of the GT sum rule (the Ikeda sum rule) up to GT GR, while the (p,n) 
reactions claim that around 90 $\%$ of the GT sum-rule strength is seen by including 
the strength beyond the GT GR up to 50 MeV \cite{wak97}. Very careful investigations of 
the GT, SD, and higher multipole strength distributions, by using charge-exchange 
reactions, are called for to see if the reduction/quenching of the $\tau \sigma$ 
strengths is partly due to the non-nucleonic ($\Delta $N$^-$) $\tau \sigma $ 
correlations \cite{eji14, boh75}. These investigations are not only interesting from
the point of view of the double beta decay but they also touch the projected double
charge-exchange reactions \cite{cap15} where the high-momentum response of nuclei is
also probed.

\section{Acknowledgments}
This work was supported by the Academy of Finland under the Finnish Center of 
Excellence Program 2012-2017 (Nuclear and Accelerator Based Program at JYFL).


\section{References}
 
\begin{thebibliography}{99}

\bibitem{eji00} H. Ejiri, Phys. Rep. 338 (2000) 265. 
\bibitem{ver12} J. Vergados, H. Ejiri, F. {\v S}imkovic, Rep. 
Prog. Phys. 75 (2012) 106301.
\bibitem{avi08} F. Avignone, S. Elliott, J. Engel, Rev. Mod. Phys. 80 (2008) 481. 
\bibitem{suh98} J. Suhonen, O. Civitarese, Phys. Rep. 300 (1998) 123.
\bibitem{hyv15} J. Hyv\"arinen and J. Suhonen, Phys. Rev. C 91 (2015) 024613.
\bibitem{fae08} A. Faessler, G.L. Fogli, E. Lisi, V. Rodin, A.M Rotunno and
F. \v Simkovic, J. Phys. G: Nucl. Part. Phys. 35 (2008) 075104.
\bibitem{suh13} J. Suhonen, O. Civitarese, Phys. Lett. B 725 (2013) 153.
\bibitem{eji14} H. Ejiri, N. Soukouti, J. Suhonen, Phys. Lett. B 729 (2014) 27.
\bibitem{suh14} J. Suhonen and  O. Civitarese, Nucl. Phys. A 924 (2014) 1.
\bibitem{eji15} H. Ejiri and J. Suhonen, J. Phys. G: Nucl. Part. Phys. 42 (2015) 055201.
\bibitem{eji78} H. Ejiri and J.I. Fujita, Phys. Rep. C 38 (1978) 85.
\bibitem{bar13} J. Barea, J. Kotila and F. Iachello, Phys. Rev. C 87 (2013) 014315.
\bibitem{bar15} J. Barea, J. Kotila and F. Iachello, Phys. Rev. C 91 (2015) 034304.
\bibitem{hol13} J.D. Holt and J. Engel, Phys. Rev. C 87 (2013) 064315.
\bibitem{vol02} C. Volpe, N. Auerbach, G. Col\`o and N. Van Giai, Phys. Rev. C 65 (2002) 044603
\bibitem{alm13} W. Almosly, E. Ydrefors, J. Suhonen, J. Phys. G:  
Nucl. Part. Phys. 40 (2013) 095201.
\bibitem{alm15} W. Almosly, E. Ydrefors, J. Suhonen, J. Phys. G:  
Nucl. Part. Phys. 42 (2015) 095106.
\bibitem{laz07} R. Lazauskas and C. Volpe, Nucl. Phys. A 792 (2007) 219.
\bibitem{haa16} M. Haaranen, P.C. Srivastava, J. Suhonen, Phys. Rev. C 93 (2016) 034308.
\bibitem{boh75} A. Bohr and  B.R. Mottelson, {\it Nuclear Structure Vol. I,II}, Benjamin,
               New York, 1969, 1975.
\bibitem{suh07} J. Suhonen, {\it From Nucleons to Nucleus: Concepts of Microscopic
Nuclear Theory}, Springer, Berlin, 2007.
\bibitem {toi98} J. Toivanen and J. Suhonen, Phys. Rev. C 57 (1998) 1237.
\bibitem {hol81} K. Holinde, Phys. Rep. 68 (1981) 121. 
\bibitem{suh88} J. Suhonen, A. Faessler, T. Taigel and  T. Tomoda, Phys. Lett. B 202 
(1988) 174 ; J. Suhonen, T. Taigel and A. Faessler, Nucl. Phys. A 486 (1988) 91.
\bibitem {aud12} G. Audi et al., Chinese Physics C 36 (2012) 1157. 
\bibitem {suh12a} J. Suhonen, O. Civitarese, J. Phys. G: Nucl. Part. 
Phys. 39 (2012) 085105.
\bibitem{bha98} M. Bhattacharya et al., Phys. Rev. C 58 (1998) 1247.
\bibitem{civ99} O. Civitarese, J. Suhonen, Nucl. Phys. A 653 (1999) 321.
\bibitem{pir15} P. Pirinen and J. Suhonen, Phys. Rev. C 91 (2015) 054309.
\bibitem {eji09} H. Ejiri, J. Phys. Soc. Jpn. 78 (2009) 074201.
\bibitem{woo90} S.E. Woosley, D.H. Hartman, D.D. Hoffman, and W.C. Haxton, 
Astrophys J. 606 (1990) 592.
\bibitem{lan03} K. Langanke and G. Mart\'\i nez-Pinedo, Rev. Mod. Phys. 75 (2003) 819.
\bibitem{che12} M.K. Cheon et al., Phys. Rev. C. 85 (2012) 065807.
\bibitem{war90} E.K. Warburton, Phys. Rev. C 42 (1990) 2479.
\bibitem{war91} E.K. Warburton, Phys. Rev. C 44 (1991) 233.
\bibitem{suz12} T. Suzuki, T. Yoshida, T. Kajino, and T. Otsuka, 
Phys. Rev. C 85 (2012) 015802.
\bibitem{zhi13} Q. Zhi, E. Caurier, J.J. Cuenca-Garc\'\i a, K. Langanke,
G. Mart\'\i nez-Pinedo, and K. Sieja, Phys. Rev. C 87 (2013) 025803.
\bibitem{wak97} T. Wakasa et al., Phys. Rev. C 55 (1997) 2909.
\bibitem{cap15} F. Cappuzzello et al., Eur. Phys. J. A 51 (2015) 145.

\end{thebibliography}
\end{document}